\documentclass[sigconf]{acmart}
\usepackage{multirow}
\usepackage{subcaption}
\usepackage{balance}
\usepackage[linesnumbered, ruled]{algorithm2e}
\AtBeginDocument{%
  }

\copyrightyear{2026}
\acmYear{2026}
\setcopyright{cc}
\setcctype{by}
\acmConference[CIKM '26]{Proceedings of the 35th ACM International Conference on Information and Knowledge Management}{November 07--11, 2026}{Rome, Italy}
\acmBooktitle{Proceedings of the 35th ACM International Conference on Information and Knowledge Management (CIKM '26), November 07--11, 2026, Rome, Italy}
\acmDOI{10.1145/3799682.3840808}
\acmISBN{979-8-4007-2539-5/2026/11}
\begin{document}

\title{Information-Guided Selective Modality-Interest Alignment for Multimodal Recommendation}

\author{Wenze Ma}
\email{mawenze991226@sjtu.edu.cn}
\affiliation{%
  \institution{Shanghai Jiao Tong University}
  \city{Shanghai}
  \country{China}
}

\author{Chenyu Sun}
\email{scy1520253537@sjtu.edu.cn}
\affiliation{%
  \institution{Shanghai Jiao Tong University}
  \city{Shanghai}
  \country{China}
}

\author{Yanmin Zhu}
\authornote{Corresponding author.}
\email{yzhu@cs.sjtu.edu.cn}
\affiliation{%
  \institution{Shanghai Jiao Tong University}
  \city{Shanghai}
  \country{China}
}

\author{Qiwen Gu}
\email{qwgu@sjtu.edu.cn}
\affiliation{%
  \institution{Shanghai Jiao Tong University}
  \city{Shanghai}
  \country{China}
}

\author{Xuhao Zhao}
\email{zhaoxuhao@sjtu.edu.cn}
\affiliation{%
  \institution{Shanghai Jiao Tong University}
  \city{Shanghai}
  \country{China}
}

\renewcommand{\shortauthors}{Wenze Ma et al.}

\begin{abstract}
Multimodal recommendation (MMRec) aims to enhance recommendation performance by leveraging rich item content from multiple modalities. However, directly incorporating all modality information does not necessarily lead to better preference modeling, since user interests are often more related to a subset of modality signals, while other signals may be weakly aligned with user preferences or even introduce noise. Although recent MMRec methods improve modality utilization through invariant learning, attention mechanisms, graph refinement, or contrastive learning, their alignment processes are often implicit or heuristic and lack a clear objective for selecting modality signals that better match user interests.
In this paper, we propose AMUR, an information-guided selective modality-interest alignment framework for multimodal recommendation. Inspired by an information-theoretic view, AMUR aims to enhance modality information that is more related to user interests while reducing the influence of less aligned signals. Specifically, AMUR first refines modality graph structures towards user behavior, and then selectively aligns shared interest-related semantics across modalities. This enables AMUR to improve modality-interest alignment while preserving useful modality-specific complementary information. Extensive experiments on three real-world datasets demonstrate the effectiveness of AMUR over competitive baselines. \textit{\href{https://github.com/Wenze1/AMUR}{The codes are available here.}}
\end{abstract}

\begin{CCSXML}
<ccs2012>
   <concept>
       <concept_id>10002951.10003317.10003347.10003350</concept_id>
       <concept_desc>Information systems~Recommender systems</concept_desc>
       <concept_significance>500</concept_significance>
       </concept>
   <concept>

 </ccs2012>
\end{CCSXML}

\ccsdesc[500]{Information systems~Recommender systems}

\keywords{multi-modal recommendation;
modality-interest alignment; information-theoretic guidance}


\maketitle

\section{Introduction}
\label{sec:introduction}

\begin{figure}[t]
    \centering
\includegraphics[width=1\linewidth]{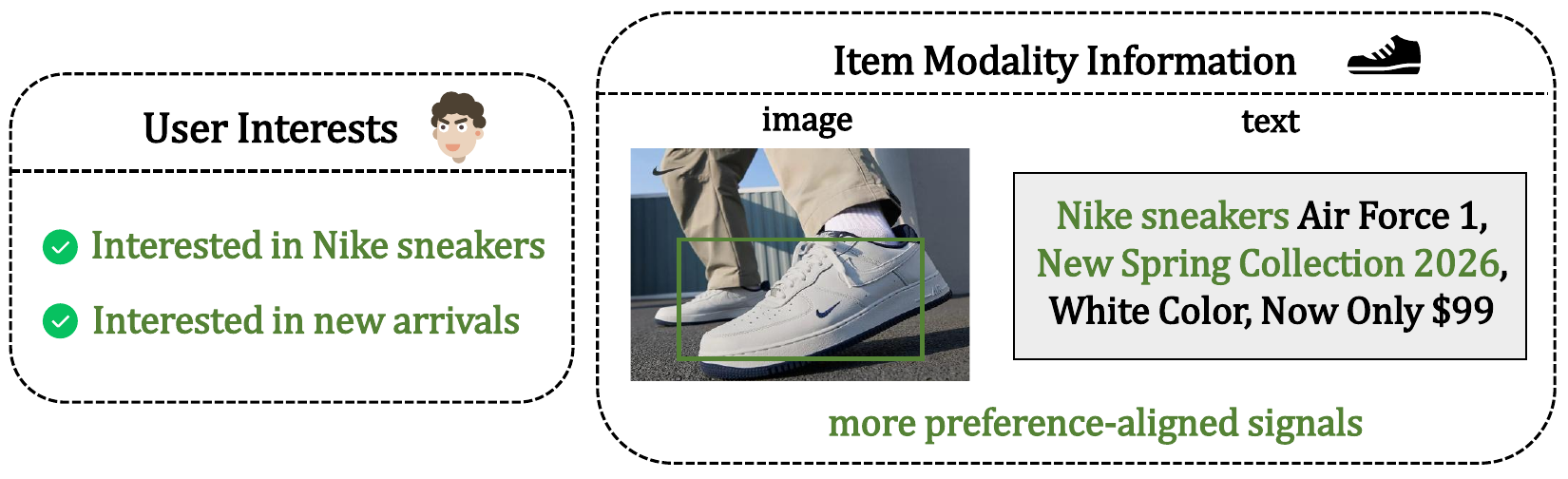}
    \caption{Illustration of modality-interest misalignment. The
highlighted green regions are more aligned with the user’s
interests, whereas other regions are less relevant to this user and may introduce preference-weak signals.}
\Description{Illustration of modality-interest misalignment between user interests and different image regions.}
    \label{fig:intro}
\end{figure}

Multimodal recommendation (MMRec) has attracted increasing attention in recent years, as items in modern recommender systems are often associated with rich auxiliary content, such as images, textual descriptions. Compared with ID-based collaborative signals, multimodal information provides fine-grained semantics about item attributes, styles, and functions, which can help alleviate data sparsity and improve user preference modeling. By incorporating such modality information, recommender systems are expected to better capture why users prefer certain items.

Existing studies have explored various ways to incorporate modality information into recommender systems, including extending matrix factorization with visual features~\cite{he2016vbpr}, propagating modality information over graph structures~\cite{wei2019mmgcn, zhang2021mining, zhou2023tale, yu2023multi}, and enhancing representation learning with self-supervised or contrastive objectives~\cite{tao2022self, zhou2023bootstrap, liu2024aligning, xu2025mentor}. These methods have demonstrated the usefulness of modality information in alleviating interaction sparsity and enriching item representations.

However, directly incorporating all modality information does not necessarily lead to better user preference modeling. In real-world recommendation scenarios, users often make decisions based on only part of the modality content. As illustrated in Figure~\ref{fig:intro}, a user may be interested in Nike sneakers and new arrivals, while other modality details, such as color or promotional information, may be less aligned with the current preference. Similarly, an item image may contain background objects or decorative details that are not useful for recommendation. If such modality signals are indiscriminately propagated, they may introduce irrelevant information into user and item representations, leading to suboptimal recommendations.

This observation highlights the importance of aligning modality
information with user interests. However, achieving such alignment is challenging, 
because the preference-aligned part of modality information is not directly observable. 
In real-world MMRec datasets, user interests are only reflected through coarse item-level interactions, such as clicks or purchases. 
These interactions indicate that a user likes an item, but do not reveal which visual regions, textual phrases, or semantic attributes actually contribute to the decision. 
As a result, models have to infer preference-aligned modality information from implicit feedback, making selective modality-interest alignment difficult in practice.

Meanwhile, the same user interest may be expressed differently across
modalities, which further complicates modality-interest alignment. For
example, a preference for a specific brand may appear as a logo in images and as a brand-related phrase in text. These modality-specific expressions introduce natural discrepancies that make the two modalities of the same item diverge in the representation space. Without modeling such cross-modal correspondence, the resulting multi-modal representations can easily create semantic gaps between modalities and become inconsistent with user interests.

Several recent methods attempt to achieve modality--interest alignment through invariant representation learning~\cite{du2022invariant}, attention mechanisms~\cite{hu2025modality}, graph refinement~\cite{zhang2021mining, yu2023multi}, or contrastive alignment~\cite{liu2024aligning, xv2024improving, xu2025mentor}. However, their alignment processes are often implicit or heuristic. In particular, they usually lack a clear objective for explaining why certain modality structures should be emphasized or suppressed. As a result, modality signals that are weakly aligned with user preferences may still be propagated or fused into the final representations.

Moreover, many alignment strategies tend to encourage global consistency across modalities. However, different modalities naturally contain both shared semantics and modality-specific complementary details. Directly aligning the full modality representations may over-suppress such modality-specific information, which can also be useful for recommendation. Therefore, an effective MMRec model should not only refine modality information toward user interests, but also perform cross-modal alignment in a selective manner, enhancing shared-interest related semantics while preserving modality-specific complementary information.

To address these issues, we propose \textbf{AMUR}, an
\textbf{information-guided selective modality-interest alignment} framework
for MMRec. The core idea of AMUR is to use an
information-theoretic view to guide the selective use of modality
information, rather than treating all modality signals as equally useful.
Specifically, we view modality information from two complementary aspects:
the mutual information part \cite{alemi2016deep} that is more aligned with user interests, and the residual part
that is weakly related to user interests. 

Based on this view, AMUR contains two key alignment stages. The first stage
is \textit{selective modality-interest refinement}. For each modality, AMUR
first builds an item graph from raw modality features, which provides a
candidate structure for modality propagation. Since visual or textual
similarity does not always match user preference, we further use behavioral
signals to calibrate this graph. Specifically, we use a KL regularizer to
make the learned edge distribution closer to the item co-occurrence patterns
observed from user interactions. In this way, edges supported by user
behavior are more likely to be retained while others are weakened. After that, an
interest-aware contrastive objective brings the refined modality
representations closer to the users, encouraging the representations to capture semantics aligned with
user interests.

The second stage is \textit{selective cross-modal shared-interest
alignment}. After each modality is refined toward user interests, AMUR
further exploits the shared-interest related semantics across modalities.
Instead of directly aligning the full visual and textual representations,
AMUR learns a dimension-wise selection gate to obtain shared-interest
subspace representations. Cross-modal consistency maximization and
discrepancy regularization are then applied only to the selected subspace.
In this way, AMUR enhances cross-modal shared semantics that are useful for
preference modeling, while preserving modality-specific complementary
information for final recommendation.

The main contributions of this work are summarized as follows:

\begin{itemize}
    \item We identify the modality-interest misalignment problem in MMRec and propose \textbf{AMUR}, an information-guided selective modality-interest alignment framework. Instead of indiscriminately using all modality signals, AMUR aims to emphasize preference-aligned modality semantics while reducing the influence of less aligned signals.

    \item We design a selective modality-interest refinement module to calibrate modality information toward user interests. It refines modality-induced item relations with behavioral signals through a KL-divergence regularizer, and enhances user-aligned modality representations via interest-aware contrastive learning.

    \item We propose a selective cross-modal shared-interest alignment module to exploit shared semantics across modalities. It learns a shared-interest subspace and applies cross-modal alignment only within the selected subspace, thereby enhancing shared-interest related semantics while preserving modality-specific complementary information.

    \item We conduct extensive experiments on three real-world MMRec datasets. The results demonstrate the effectiveness of AMUR compared to competitive baselines.
\end{itemize}

\section{Methodology}
\label{sec:method}

\begin{figure*}[t]
    \centering
\includegraphics[width=0.95\linewidth]{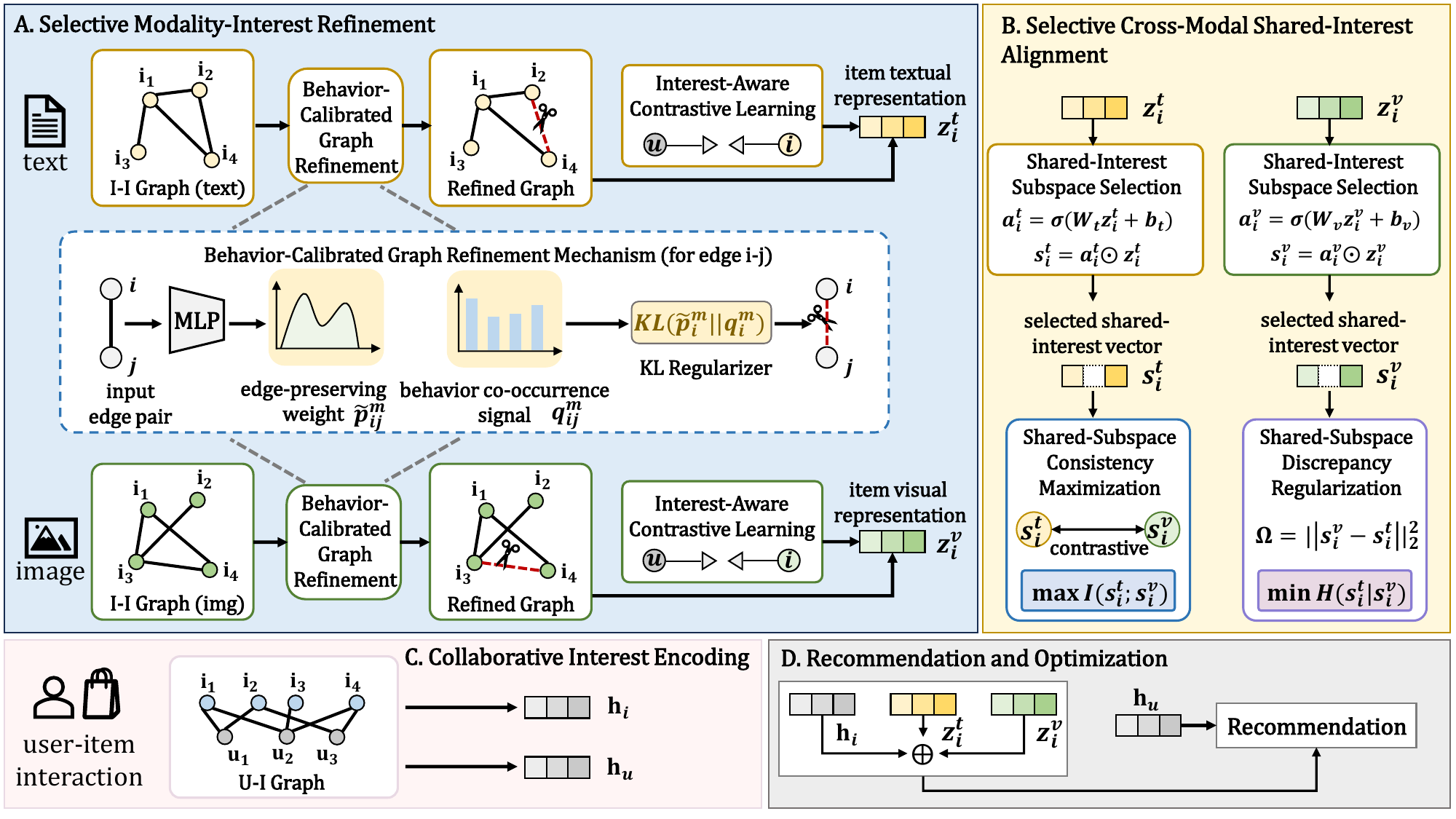}
    \caption{The overall framework of AMUR. AMUR first encodes collaborative interests from user--item interactions, then performs selective modality-interest refinement to obtain user-aligned modality representations. It further conducts selective cross-modal shared-interest alignment, and finally integrates collaborative and modality representations for recommendation.}
    \Description{A framework diagram of AMUR. User--item interactions are first encoded into collaborative representations. Modality-specific features are then refined according to user interests, followed by cross-modal shared-interest alignment. The resulting collaborative and modality representations are fused and used for recommendation.}
    \label{fig:model}
\end{figure*}

\subsection{Problem Formulation}
% free id emb-->u i
% modality feat--> iv it
Let $\mathcal{U}$ and $\mathcal{I}$ be the user set and the item set. The user-item interaction matrix is denoted as $\boldsymbol{\mathrm{R}} \in \mathbb{R}^{|\mathcal{U}|\times|\mathcal{I}|}$, where $\boldsymbol{\mathrm{R}}_{ui}=1$ indicates that user $u$ has interacted with item $i$, and $\boldsymbol{\mathrm{R}}_{ui}=0$ otherwise. For each user $u$ and item $i$, we have their id embeddings $e_{u} ^{id}, e_{i}^{id} \in \mathbb{R}^d$ from embedding layers $\boldsymbol{\mathrm{E}}_{\mathrm{U}}^{id}$ and  $\boldsymbol{\mathrm{E}}_{\mathrm{I}}^{id}$. Under MMRec settings, each item is associated with modality features $e^m _i\in \mathbb{R}^{d_m}$, where $m\in \{v,t\}$, including visual feature $e_i^v$ and textual feature $e_i^t$.

The goal of MMRec is to predict the preference score $\hat{y}_{ui}$:
\begin{equation}
    \hat{y}_{ui} = f\left(u, i \mid e_{u}^{id}, e_{i}^{id}, e_{i}^{v}, e_{i}^{t}, \boldsymbol{\mathrm{R}}\right),
\end{equation}
and recommend a top-$k$ list of items that $u$ might be interested in. In this work, rather than indiscriminately using all modality information,
we focus on information-guided selective modality alignment, which aims to
preserve preference-relevant modality semantics while reducing the influence
of modality signals that are weakly related to user interests.

\subsection{Overview}

The overall framework of AMUR is illustrated in Figure~\ref{fig:model}. 
AMUR aims to selectively align item modality information with user interests 
under an information-guided view. Instead of directly using all modality 
signals, AMUR first extracts collaborative interests from user--item 
interactions, and then uses them to guide modality refinement.

Specifically, AMUR contains two key alignment stages. The first stage performs
selective modality-interest refinement, where raw modality graphs are refined
by behavior-calibrated signals and the learned modality representations are
pulled closer to user interests. The second stage performs selective
cross-modal shared-interest alignment, where only the selected shared-interest
subspace is aligned across visual and textual modalities. This avoids forcing
all modality-specific information to be identical while still enhancing
cross-modal shared semantics.

Finally, AMUR integrates collaborative representations and refined modality
representations for recommendation, and jointly optimizes all objectives.

\subsection{Collaborative Interest Encoding} \label{collab}

User-item interactions provide direct evidence of user preferences.
Therefore, we first encode collaborative signals from the interaction
graph and use them as preference anchors for subsequent modality
refinement. Specifically, we construct a user-item graph based on the
interaction matrix $\boldsymbol{\mathrm{R}}$:
\begin{equation}
\boldsymbol{\mathrm{A}}=
\begin{bmatrix}
  0& \boldsymbol{\mathrm{R}} \\
  \boldsymbol{\mathrm{R}}^{\mathrm{T}}& 0
\end{bmatrix}.
\end{equation}
Following LightGCN~\cite{he2020lightgcn}, we normalize
$\boldsymbol{\mathrm{A}}$ as $\hat{\boldsymbol{\mathrm{A}}}$ and propagate
ID embeddings
$\boldsymbol{\mathrm{H}}^0=
\begin{bmatrix}
 \boldsymbol{\mathrm{E}}_{\mathrm{U}}^{id}\\
 \boldsymbol{\mathrm{E}}_{\mathrm{I}}^{id}
\end{bmatrix}$
over the interaction graph:
\begin{equation}
    \boldsymbol{\mathrm{H}}^{l+1}= \hat{\boldsymbol{\mathrm{A}}}\boldsymbol{\mathrm{H}}^{l},\quad l\in\{0, \cdots, L-1\},
\end{equation}
\begin{equation}
        \boldsymbol{\mathrm{H}}= \dfrac{1}{L+1}\sum_{l=0}^{L}\boldsymbol{\mathrm{H}}^{l}.
\end{equation}
After propagation, we obtain the user collaborative representation
$\boldsymbol{\mathrm{h}}_u$ and item collaborative representation
$\boldsymbol{\mathrm{h}}_i$. These representations summarize high-order
behavioral patterns from observed interactions and serve as preference
anchors for calibrating modality representations toward user interests.

\subsection{Selective Modality-Interest Refinement}
As discussed above, user interests often focus on only a subset of item modality contents. Therefore, directly propagating or aligning all modality information may inject preference-irrelevant signals into user preference modeling \cite{zhang2021mining}. 

To alleviate this issue, existing MMRec methods often learn invariant representations across modalities \cite{du2022invariant}, incorporate attention mechanisms \cite{hu2025modality} or refining modality representations and structures
\cite{zhang2021mining, liu2024aligning, xu2025mentor}. Although effective, these
methods usually rely on implicit or heuristic strategies, making it
difficult to interpret which part of modality information is encouraged to
match user interests and which part should be suppressed.

In this paper, we use an information-guided view to motivate a selective
modality-interest refinement process.

\subsubsection{Information-Guided Motivation}

Ideally, a desirable modality representation should preserve information
that is useful for explaining user preferences, while reducing the influence
of modality details that are weakly related to user interests. We describe
this goal from an information-theoretic perspective.

For each modality \(m \in \{v,t\}\), let \(\boldsymbol{z}^m_i \in
\mathbb{R}^d\) denote the modality representation of item \(i\)
encoded\footnote{We will illustrate details of the encoder in
Equation~(\ref{itemrep}).} from raw modality features. Let
\(\boldsymbol{\mathrm{h}}_u\) denote the collaborative representation of user \(u\)
learned from historical interactions. The entropy of
\(\boldsymbol{z}^m_i\) can be decomposed as:
\begin{equation}
H(\boldsymbol{z}_i^m)
=
\underset{\text{preference-relevant part}}{\underbrace{
I(\boldsymbol{z}_i^m;\boldsymbol{\mathrm{h}}_u)}}
+
\underset{\text{unexplained residual part}}{\underbrace{
H(\boldsymbol{z}_i^m|\boldsymbol{\mathrm{h}}_u)}} .
\end{equation}

This decomposition provides an intuitive view of modality-interest
refinement. 
\begin{itemize}
    \item The mutual information term \(I(\boldsymbol{z}_i^m;\boldsymbol{\mathrm{h}}_u)\) measures how much information in the modality representation is related to the user preference representation. A larger value means that the modality representation contains more preference-relevant semantics.
    \item The conditional entropy term \(H(\boldsymbol{z}_i^m|\boldsymbol{\mathrm{h}}_u)\) measures the remaining uncertainty in the modality representation after the user preference representation is known. In our scenario, this residual part may include modality details that are weakly related or irrelevant to the current user interest.
\end{itemize}

Based on this view, AMUR has two optimization directions: increasing the
preference-relevant part by enhancing
\(I(\boldsymbol{z}_i^m;\boldsymbol{\mathrm{h}}_u)\), and reducing the influence of
the unexplained residual part associated with
\(H(\boldsymbol{z}_i^m|\boldsymbol{\mathrm{h}}_u)\). However, directly estimating
mutual information and conditional entropy is generally intractable in
large-scale recommendation scenarios~\cite{alemi2016deep}. Therefore, we
instantiate this information-guided objective with two tractable modules:
a behavior-calibrated graph refinement module to suppress
noisy modality structures, and an interest-aware
contrastive learning module to enhance user-aligned modality semantics.

\subsubsection{Modality-Initialized Graph Construction}

Following ~\cite{su2024soil, xu2025mentor, yu2023multi}, we first construct a modality-initialized item--item graph as the candidate structure for each modality. Specifically, for each modality \(m \in \{v,t\}\), we compute the cosine similarity between item \(i\) and item \(j\) based on their raw modality features \(\boldsymbol{e}_i^m\) and \(\boldsymbol{e}_j^m\):
\begin{equation}
    o_{i,j}^m =
    \frac{(\boldsymbol{e}_i^m)^{\mathrm{T}}\boldsymbol{e}_j^m}
    {\|\boldsymbol{e}_i^m\| \cdot \|\boldsymbol{e}_j^m\|}.
\end{equation}
Then, each item is connected to its top-\(K\) most similar items under modality \(m\):
\begin{equation}
s_{i,j}^m =
\begin{cases}
1, & o_{i,j}^m \in \mathrm{TopK}(\{o_{i,k}^m \mid k \in \mathcal{I}\}),\\
0, & \text{otherwise},
\end{cases}
\end{equation}
where \(s_{i,j}^m\) denotes the entry at the \(i\)-th row and \(j\)-th column of the modality-initialized item--item graph
\(\boldsymbol{S}^m \in \mathbb{R}^{|\mathcal{I}|\times|\mathcal{I}|}\).  It provides a lightweight structural prior that organizes items into local neighborhoods based on their modality features, facilitating subsequent feature extraction. Such a graph learning approach is widely adopted by previous work, which has shown effectiveness in MMRec models.

\subsubsection{Behavior-Calibrated Graph Refinement}

The modality-initialized graph \(\boldsymbol{S}^m\) captures item similarity
in the modality feature space, but such similarity does not necessarily
reflect user-interest relations. For example, two smartphones may appear
visually similar, but appeal to different user groups. 
Therefore, directly
propagating over the raw modality graph may introduce
preference-inconsistent signals. To address this issue, we refine
\(\boldsymbol{S}^m\) by learning which modality edges should be preserved
under the guidance of collaborative behavior.

For each candidate edge \((i,j)\) in \(\boldsymbol{S}^m\), we first estimate
an edge-preservation probability:
\begin{equation}
    p_{ij}^m=
    \sigma\left(
    \operatorname{MLP}_{\theta}^m
    ([e_i^{id},e_j^{id},o_{ij}^m])
    \right),
\end{equation}
where \([e_i^{id},e_j^{id},o_{ij}^m]\) denotes
the concatenation of the ID embeddings of items \(i\) and \(j\), together
with their raw modality similarity \(o_{ij}^m\). The MLP maps the input to
a scalar score, and the sigmoid function converts it into
\(p_{ij}^m\in(0,1)\). A larger \(p_{ij}^m\) indicates that edge \((i,j)\)
is more likely to be useful for preference-aware modality propagation.

Based on \(p_{ij}^m\), we sample a binary edge mask:
\begin{equation}
    \beta_{ij}^m\sim \operatorname{Bern}(p_{ij}^m),
\end{equation}
where \(\operatorname{Bern}(p_{ij}^m)\) denotes a Bernoulli distribution
with success probability \(p_{ij}^m\). The sampled value \(\beta_{ij}^m\)
acts as an edge switch: \(\beta_{ij}^m=1\) means that the edge is preserved,
while \(\beta_{ij}^m=0\) means that the edge is filtered out. Here we adopt the Gumbel-softmax trick \cite{jang2016categorical} to obtain differentiable binary masks. After sampling
masks for all candidate edges, the refined modality graph is obtained by:
\begin{equation}
    \hat{\boldsymbol{S}}^m=\boldsymbol{\beta}^m\odot \boldsymbol{S}^m,
\end{equation}
where \(\odot\) denotes the Hadamard product. Thus,
\(\hat{\boldsymbol{S}}^m\) is a preference-calibrated subgraph of
\(\boldsymbol{S}^m\), where more preference-consistent modality edges are
more likely to be retained. 

To further guide the refinement process, we introduce behavioral
co-occurrence as a preference-aware reference. Specifically, we compute:
\begin{equation}
    f_{ij}=[\boldsymbol{R}^{\top}\boldsymbol{R}]_{ij},
\end{equation}
where \(f_{ij}\) counts how many users have interacted with both items
\(i\) and \(j\). A larger \(f_{ij}\) indicates that the two items are more
likely to be close from the perspective of collaborative preference. 

We then convert the behavior co-occurrence signal into a reference
distribution over the candidate modality neighbors :
\begin{equation}
    q_{ij}^m=
    \frac{s_{ij}^m f_{ij}}
    {\sum_{k}s_{ik}^m f_{ik}+\epsilon},
\end{equation}
where \(s_{ij}^m\) ensures that the reference distribution is defined only
over modality-derived candidate edges, and \(\epsilon\) is a small constant
for numerical stability.

Similarly, we normalize the learned edge-preservation probabilities over
the candidate neighbors:
\begin{equation}
    \tilde{p}_{ij}^m
    =
    \frac{s_{ij}^m p_{ij}^m}
    {\sum_{k}s_{ik}^m p_{ik}^m+\epsilon}.
\end{equation}

We then regularize the learned modality-neighbor distribution toward the
behavior-informed reference distribution by minimizing their KL divergence:
\begin{equation}
\begin{aligned}
    \mathcal{L}_{br}^m
    &=
    \sum_i
    D_{\mathrm{KL}}
    \left(
    \tilde{\boldsymbol{p}}_i^m
    \|
    \boldsymbol{q}_i^m
    \right)  \\
    &=
    \sum_i \sum_j
    \tilde{p}_{ij}^m
    \log
    \frac{\tilde{p}_{ij}^m}
    {q_{ij}^m+\epsilon} .
\end{aligned}
\end{equation}

From the information-guided view, this surrogate discourages modality
structures that are weakly supported by user preference signals, thereby
reducing the influence of the unexplained residual part associated with
\(H(\boldsymbol{z}_i^m|\boldsymbol{\mathrm{h}}_u)\).

\subsubsection{Interest-Aware Contrastive Learning}

After obtaining the refined modality graph \(\hat{\boldsymbol{S}}^m\), we further extract modality representations and align them with user interests. The graph refinement module suppresses preference-inconsistent modality relations, while this module enhances the preference-relevant part by pulling modality representations closer to the collaborative interests of users who have interacted with them.

Specifically, we encode the modality representation of items by propagating modality features over the refined graph:
\begin{equation}
\label{itemrep}
    \boldsymbol{Z}^m
    =
    \hat{\boldsymbol{S}}^m
    \left(
    \boldsymbol{\mathrm{E}}_{\mathrm{I}}^{id}
    \odot
    \operatorname{MLP}_{\gamma}^m(\boldsymbol{\mathrm{E}}^m)
    \right),
\end{equation}
where
\(\boldsymbol{\mathrm{E}}^m\) denotes raw modality features under modality \(m\),
\(\operatorname{MLP}_{\gamma}^m(\cdot)\) projects raw modality features into the latent space, and \(\odot\) denotes element-wise multiplication. We use \(\boldsymbol{z}_i^m\) to denote the \(i\)-th row of \(\boldsymbol{Z}^m\).

Then, for each observed user--item interaction \((u,i)\), we regard \((\boldsymbol{\mathrm{h}}_u,\boldsymbol{z}_i^m)\) as a positive pair, since item \(i\) reflects the preference of user \(u\) in historical behavior. Other items in the batch are treated as negative samples. The interest-aware contrastive loss is defined as:
\begin{equation}
    \mathcal{L}_{cl}^m
    =
    -\frac{1}{|\mathcal{B}|}
    \sum_{(u,i)\in \mathcal{B}}
    \log
    \frac{
    \exp\left(s(\boldsymbol{\mathrm{h}}_u,\boldsymbol{z}_i^m)/\tau\right)
    }
    {
    \sum_{j\in \mathcal{B}_{\mathcal{I}}}
    \exp\left(s(\boldsymbol{\mathrm{h}}_u,\boldsymbol{z}_j^m)/\tau\right)
    },
\end{equation}
where \(\mathcal{B}\) denotes a mini-batch of observed interactions,
\(\mathcal{B}_{\mathcal{I}}\) denotes the item set in the batch,
\(s(\cdot,\cdot)\) is the cosine similarity function, and \(\tau\) is the temperature parameter.

This contrastive objective encourages the modality representation of the interacted item to be closer to the user collaborative representation than other items. From the information-guided view, it serves as a tractable surrogate for enhancing the preference-relevant term \(I(\boldsymbol{z}_i^m;\boldsymbol{\mathrm{h}}_u)\). Following the standard InfoNCE formulation \cite{oord2018representation}, we have:
\begin{equation}
    I(\boldsymbol{z}_i^m;\boldsymbol{\mathrm{h}}_u)
    \geq
    \log |\mathcal{B}_{\mathcal{I}}|
    -
    \mathcal{L}_{cl}^m .
\end{equation}
Therefore, minimizing \(\mathcal{L}_{cl}^m\) maximizes a lower-bound surrogate of the mutual information between modality representations and user interests. This helps the learned modality representations preserve user-aligned semantics after behavior-calibrated graph refinement.

Combining the behavior-calibrated graph refinement and interest-aware contrastive learning, the objective of modality-interest refinement is:
\begin{equation}
    \mathcal{L}_{mir}
    =
    \sum_{m\in\{v,t\}}
    \left(
    \mathcal{L}_{br}^m
    +
    \mathcal{L}_{cl}^m
    \right).
\end{equation}

\subsection{Selective Cross-Modal Shared-Interest Alignment}

After modality-interest refinement, each modality representation
\(\boldsymbol{z}_i^m\) has been calibrated toward user interests.
However, different modalities may still express the same interest-related
semantics in different forms. For example, a user's preference for a brand
may appear as a logo in the visual modality and as a brand name in the
textual modality. Such cross-modal shared semantics can further improve
recommendation robustness.

A straightforward solution is to directly align the full visual and textual
representations. However, this may over-suppress modality-specific
complementary information, since visual and textual modalities naturally
contain different but useful signals. Therefore, instead of enforcing
full-space alignment, we propose a selective shared-interest alignment
module. The key idea is to first select the dimensions that are more suitable
for cross-modal shared semantic alignment, and then apply cross-modal
objectives only within the selected shared-interest subspace.
\subsubsection{Shared-Interest Subspace Selection}

For each modality representation \(\boldsymbol{z}_i^m\), we learn a
dimension-wise soft selection gate:
\begin{equation}
    \boldsymbol{a}_i^m
    =
    \sigma
    \left(
    \boldsymbol{\mathrm{W}}_m \boldsymbol{z}_i^m
    +
    \boldsymbol{\mathrm{b}}_m
    \right),
\end{equation}
where \(\boldsymbol{a}_i^m \in (0,1)^d\) denotes the selection weights for
different dimensions, and \(\sigma(\cdot)\) is the sigmoid function. Then,
we obtain the selected shared space by:
\begin{equation}
    \boldsymbol{s}_i^m
    =
    \boldsymbol{a}_i^m
    \odot
    \boldsymbol{z}_i^m,
\end{equation}
where \(\odot\) denotes element-wise multiplication.

The gate \(\boldsymbol{a}_i^m\) softly reweights the modality representation
and highlights dimensions that are more suitable for cross-modal shared
semantic alignment. In this way, cross-modal objectives are not directly
imposed on the entire modality representation \(\boldsymbol{z}_i^m\), but
only on the selected shared space \(\boldsymbol{s}_i^m\).
This design helps enhance shared semantics across modalities while reducing
the risk of forcing modality-specific complementary information to be
identical.
\subsubsection{Information-Guided Shared-Subspace Alignment}

From an information-guided view, the selected representation
\(\boldsymbol{s}_i^m\) should contain shared semantics that can be explained
by another modality, while reducing unnecessary discrepancy in the selected
subspace. For two different modalities \(m,m'\in\{v,t\}\), the information
contained in \(\boldsymbol{s}_i^m\) can be written as:
\begin{equation}
H(\boldsymbol{s}_i^m)
=
\underset{\text{shared semantic consistency}}{\underbrace{
I(\boldsymbol{s}_i^m;\boldsymbol{s}_i^{m'})}}
+
\underset{\text{shared-subspace discrepancy}}{\underbrace{
H(\boldsymbol{s}_i^m|\boldsymbol{s}_i^{m'})}} .
\end{equation}

This decomposition provides two optimization directions. First, the mutual
information term \(I(\boldsymbol{s}_i^m;\boldsymbol{s}_i^{m'})\) should be
enhanced, so that the selected subspaces from different modalities preserve
consistent item semantics. Second, the conditional entropy term
\(H(\boldsymbol{s}_i^m|\boldsymbol{s}_i^{m'})\) should be reduced, so that
unnecessary discrepancy within the selected shared-interest subspace can be
regularized.
\subsubsection{Shared-Subspace Consistency Maximization}

To enhance cross-modal shared semantics, we maximize the agreement between
the selected representations of the same item. Specifically, for item \(i\),
we treat \((\boldsymbol{s}_i^v,\boldsymbol{s}_i^t)\) as a positive pair,
and \((\boldsymbol{s}_i^v,\boldsymbol{s}_j^t)\) with \(j\neq i\) as negative
pairs. The shared-subspace contrastive loss is defined as:
\begin{equation}
    \mathcal{L}_{sc}
    =
    -\frac{1}{|\mathcal{B}_{\mathcal{I}}|}
    \sum_{i\in\mathcal{B}_{\mathcal{I}}}
    \log
    \frac{
    \exp\left(
    s(\boldsymbol{s}_i^v,\boldsymbol{s}_i^t)/\tau
    \right)
    }
    {
    \sum_{j\in\mathcal{B}_{\mathcal{I}}}
    \exp\left(
    s(\boldsymbol{s}_i^v,\boldsymbol{s}_j^t)/\tau
    \right)
    },
\end{equation}
where \(\mathcal{B}_{\mathcal{I}}\) denotes the item set in a mini-batch,
\(s(\cdot,\cdot)\) is the cosine similarity function, and \(\tau\) is the
temperature parameter.

Following the standard InfoNCE formulation, minimizing
\(\mathcal{L}_{sc}\) maximizes a lower-bound surrogate of
\(I(\boldsymbol{s}_i^v;\boldsymbol{s}_i^t)\). Therefore, this objective
encourages the selected visual and textual subspaces to capture consistent
and discriminative shared semantics.
\subsubsection{Shared-Subspace Discrepancy Regularization}

Besides maximizing shared semantic consistency, we further regularize the
discrepancy between selected subspaces. Directly estimating
\(H(\boldsymbol{s}_i^m|\boldsymbol{s}_i^{m'})\) is intractable. Following a
variational view \cite{kingma2013auto}, we introduce a conditional distribution
\(\Omega(\boldsymbol{s}_i^m|\boldsymbol{s}_i^{m'})\). Then we have:
\begin{equation}
\begin{aligned}
H(\boldsymbol{s}_i^m|\boldsymbol{s}_i^{m'})
&\leq
-\mathbb{E}_{P(\boldsymbol{s}_i^m,\boldsymbol{s}_i^{m'})}
\log
\Omega(\boldsymbol{s}_i^m|\boldsymbol{s}_i^{m'}).
\end{aligned}
\end{equation}

We instantiate \(\Omega(\boldsymbol{s}_i^m|\boldsymbol{s}_i^{m'})\) as an
isotropic Gaussian distribution:
\begin{equation}
    \Omega(\boldsymbol{s}_i^m|\boldsymbol{s}_i^{m'})
    =
    \mathcal{N}
    \left(
    \boldsymbol{s}_i^m
    \mid
    \boldsymbol{s}_i^{m'},
    \gamma \boldsymbol{\mathrm{I}}
    \right),
\end{equation}
where \(\gamma\boldsymbol{\mathrm{I}}\) is the covariance matrix. Under this
instantiation, minimizing the negative log-likelihood is equivalent to
minimizing the squared distance between the selected subspaces. Therefore,
we define the shared-subspace discrepancy regularization as:
\begin{equation}
    \mathcal{L}_{sd}
    =
    \sum_{i\in\mathcal{B}_{\mathcal{I}}}
    \left\|
    \boldsymbol{s}_i^v
    -
    \boldsymbol{s}_i^t
    \right\|_2^2 .
\end{equation}

This term reduces unnecessary discrepancy only in the selected
shared-interest subspace, rather than forcing the full visual and textual
representations to be identical.

The objective of selective cross-modal shared-interest alignment is:
\begin{equation}
    \mathcal{L}_{sia}
    =
    \mathcal{L}_{sc}
    +
    \mathcal{L}_{sd}.
\end{equation}

Although related, these two terms play \emph{complementary} roles in aligning semantics.
\begin{itemize}
    \item The contrastive term $\mathcal{L}_{sc}$
    imposes \textit{discriminative} alignment by separating mismatched modality pairs, preventing trivial solutions (e.g., all modalities collapsing to the same point) and preserving item-level distinctions.
    \item The discrepancy term $\mathcal{L}_{sd}$
    enforces \textit{local alignment} by penalizing systematic deviation between modalities of the same item, ensuring that different views collapse toward a common semantic anchor. 
\end{itemize}
By applying cross-modal objectives on \(\boldsymbol{s}_i^v\) and
\(\boldsymbol{s}_i^t\), AMUR selectively enhances shared-interest related
semantics across modalities while preserving modality-specific information
in the original representations \(\boldsymbol{z}_i^v\) and
\(\boldsymbol{z}_i^t\) for final recommendation.

\subsection{Recommendation and Optimization}

After obtaining the collaborative representations and modality representations,
we integrate them for final recommendation. Specifically, for each item \(i\),
we combine its collaborative representation \(\boldsymbol{\mathrm{h}}_i\) and
the refined modality representations \(\boldsymbol{z}_i^v\) and
\(\boldsymbol{z}_i^t\) as:
\begin{equation}
    \boldsymbol{\mathrm{h}}_i^{*}
    =
    \boldsymbol{\mathrm{h}}_i
    +
    \boldsymbol{z}_i^v
    +
    \boldsymbol{z}_i^t .
\end{equation}

It is worth noting that the selected shared-interest representations
\(\boldsymbol{s}_i^v\) and \(\boldsymbol{s}_i^t\) are only used for
cross-modal shared-interest alignment. The final recommendation still uses
the complete modality representations \(\boldsymbol{z}_i^v\) and
\(\boldsymbol{z}_i^t\), so that modality-specific complementary information
can be preserved.

Then, the preference score between user \(u\) and item \(i\) is calculated by:
\begin{equation}
    \hat{y}_{ui}
    =
    \boldsymbol{\mathrm{h}}_u^{\mathrm{T}}
    \boldsymbol{\mathrm{h}}_i^{*}.
\end{equation}

We adopt the Bayesian Personalized Ranking (BPR) loss \cite{rendle2012bpr} for recommendation:
\begin{equation}
    \mathcal{L}_{bpr}
    =
    -\sum_{(u,i,j)\in \mathcal{D}}
    \log
    \sigma
    \left(
    \hat{y}_{ui}
    -
    \hat{y}_{uj}
    \right),
\end{equation}
where \((u,i,j)\) denotes a training triplet, in which item \(i\) is an
observed positive item for user \(u\), and item \(j\) is a sampled negative
item.

Finally, AMUR is optimized by jointly considering recommendation loss,
modality-interest refinement loss, and selective shared-interest alignment
loss:
\begin{equation}
    \mathcal{L}
    =
    \mathcal{L}_{bpr}
    +
    \alpha \mathcal{L}_{mir}
    +
    \beta \mathcal{L}_{sia} ,
\end{equation}
where \(\mathcal{L}_{mir}\) denotes the objective of selective
modality-interest refinement, \(\mathcal{L}_{sia}\) denotes the objective
of selective cross-modal shared-interest alignment, \(\alpha\) and
\(\beta\) are trade-off hyperparameters.

\begin{table}[t]
    \caption{Statistics of three widely-used multi-modal recommendation datasets.}
    \label{tab:stat}
    \centering
    % \resizebox{\textwidth}{!}{
    \scalebox{0.9}{
    \begin{tabular}{c|ccc}
    \toprule
    Dataset & \textbf{Baby} & \textbf{Sports} & \textbf{Clothing} \\
     \midrule
    Modality & V\;\space\space\,\, T  & V\;\space\space\,\, T &V\space\space\space\space T\\ 
    Embed Dim  & 4096 \space384 & 4096 \,384 & 4096 384 \\ 
    User& 19445  & 35598 & 39387  \\
    Item & 7050  & 18357  & 23033   \\ 
    Interactions & 160792 & 296337 & 278677 \\ 
    \bottomrule
    \end{tabular}
    }
\end{table}

\section{Experiments}
\subsection{Experimental Setup}
\subsubsection{Datasets}
Following prior multi-modal recommendation studies 
\cite{zhou2023tale, yu2023multi, guo2024lgmrec, zhou2023mmrec}, 
we evaluate our model on three widely-adopted benchmark datasets: 
\textit{Baby}, \textit{Sports}, and \textit{Clothing}. 
These datasets contain rich multi-modal information, including 
item images, textual descriptions, and user–item interaction records, 
and have been extensively used to assess multi-modal representation 
learning in recommendation systems. They span three distinct
domains with heterogeneous visual and textual characteristics, 
providing diverse scenarios for evaluating multi-modal alignment.

To ensure fair comparison and reproducibility, we follow the standard 
preprocessing and training protocols used in previous work 
\cite{zhou2023mmrec, zhou2023comprehensive}. 
Table~\ref{tab:stat} summarizes the dataset statistics.

\subsubsection{Baseline Methods and Evaluation Metrics}

We compare AMUR with representative MMRec baselines from different categories: 
\textbf{MF-based method}: VBPR~\cite{he2016vbpr}; 
\textbf{invariant and causal learning-based methods}: InvRL~\cite{du2022invariant} and ISOLATOR~\cite{xv2025unveiling}; 
\textbf{graph and topology-based methods}: MMGCN~\cite{wei2019mmgcn}, LATTICE~\cite{zhang2021mining}, FREEDOM~\cite{zhou2023tale}, MGCN~\cite{yu2023multi}, LGMRec~\cite{guo2024lgmrec}, DA-MRS~\cite{xv2024improving}, and TMLP~\cite{huang2025beyond}; 
\textbf{SSL and alignment-based methods}: SLMRec~\cite{tao2022self}, BM3~\cite{zhou2023bootstrap}, and MENTOR~\cite{xu2025mentor};
\textbf{diffusion-based methods}: DiffMM~\cite{jiang2024diffmm} and CCDRec~\cite{yang2025curriculum}; 
and \textbf{Transformer/MLLM-based methods}: MIG-GT~\cite{hu2025modality} and BeFA~\cite{fan2025befa}. 
We adopt Recall@\(K\) and NDCG@\(K\) with \(K=\{10,20\}\) as evaluation metrics.

\subsubsection{Parameter Settings}

Following~\cite{zhou2023mmrec}, we fix the batch size to 2048, the embedding dimension \(d\) to 64, and the learning rate to 0.001 for all methods. For each baseline, we tune the key hyperparameters according to the ranges suggested in the original papers or released codes. For AMUR, we search the coefficient \(\alpha\) of selective modality-interest refinement and the coefficient \(\beta\) of selective cross-modal shared-interest alignment in \(\{0, 0.001, 0.01, 0.1\}\). The influence of these two coefficients is further analyzed in the hyperparameter study.

\subsection{Overall Performance}\label{pc}
\begin{table*}[t]
    \caption{Performance comparisons of different approaches over three datasets. The best results of each metric are marked in bold, while the best results of baselines is underlined. * represents that the improvements of AMUR compared with best baselines are statistically significant for $p<0.05$.}
    \label{tab:comparison}
    \centering
    \resizebox{\textwidth}{!}{
    \begin{tabular}{c|cccc|cccc|cccc}
    \toprule
    \multirow{2}{*}{\textbf{Models}} &
    \multicolumn{4}{c|}{\textbf{Baby}} &  \multicolumn{4}{c|}{\textbf{Sports}}  & \multicolumn{4}{c}{\textbf{Clothing}} \\
    & \textbf{R@10} & \textbf{R@20} & \textbf{N@10} & \textbf{N@20} & \textbf{R@10} & \textbf{R@20} & \textbf{N@10} & \textbf{N@20} & \textbf{R@10} & \textbf{R@20} & \textbf{N@10} & \textbf{N@20} \\
    \midrule
    \textbf{VBPR} & 0.0424 & 0.0664 & 0.0223 & 0.0285 & 0.0559& 0.0858& 0.0307& 0.0384 & 0.0281	& 0.0410 & 0.0157& 0.0190\\
    \midrule
    \textbf{InvRL} & 0.0320 & 0.0519 & 0.0171 & 0.0222 & 0.0324 & 0.0504 & 0.0176 & 0.0223 & 0.0227 & 0.0363 & 0.0122 & 0.0157 \\
        \textbf{ISOLATOR} & 0.0628 & 0.0957 & 0.0346 & 0.0430 & 0.0730 & 0.1112 & 0.0395 & 0.0493 & 0.0659 & 0.0972 & 0.0358 & 0.0437 \\
    \midrule
    \textbf{MMGCN} & 0.0397&0.0644&0.0209&0.0272& 0.0381&0.0615&0.0200&0.0260&0.0221&0.0362&0.0113&0.0149\\
    \textbf{LATTICE} &0.0536&0.0858&0.0287&0.0370&0.0618&0.0950&0.0337&0.0423&0.0459&0.0702&0.0253&0.0306\\
    \textbf{FREEDOM} & 0.0619&0.0987&0.0326&0.0420&0.0711&0.1090&0.0385&0.0482&0.0639&0.0940&0.0344&0.0421\\

    \textbf{MGCN} & 0.0614&0.0963&0.0327&0.0417&0.0732&0.1105&0.0397&0.0493&0.0659&0.0967&0.0360&\underline{0.0438}\\
    \textbf{LGMRec} &0.0650&0.0989&0.0349&0.0436&0.0723&0.1090&0.0392&0.0487&0.0552&0.0825&0.0301&0.0371\\
    \textbf{DA-MRS}&0.0617&0.0960&0.0337&0.0425&0.0706&0.1073&0.0381&0.0476&0.0603&0.0902&0.0328&0.0404 \\
    \textbf{TMLP} &0.0658&\underline{0.1009}&\underline{0.0356}&\underline{0.0447}& \underline{0.0751}&\underline{0.1137}&0.0402&\underline{0.0501}&0.0654&0.0957&\underline{0.0360}&0.0437\\
    \midrule
            \textbf{SLMRec} & 0.0540&0.0810&0.0285&0.0357 &0.0676&0.1017&0.0374&0.0462 & 0.0452&0.0675&0.0247&0.0303\\
\textbf{BM3}&0.0547&0.0866&0.0290&0.0373&0.0646&0.0981&0.0353&0.0439&0.0429&0.0630&0.0232&0.0283\\
\textbf{MENTOR}&0.0616&0.0995&0.0336&0.0433&0.0718&0.1092&0.0389&0.0485&0.0606&0.0890&0.0325&0.0397\\
    \midrule
        \textbf{DiffMM} &0.0591&0.0924&0.0320&0.0405&0.0690&0.1049&0.0371&0.0463&0.0569&0.0897&0.0313&0.0397\\
        \textbf{CCDRec} &\underline{0.0661}&0.1006&0.0352&0.0441&0.0730&0.1108&0.0392&0.0489
&\underline{0.0661}&\underline{0.0975}&0.0358&\underline{0.0438}\\
    \midrule
        \textbf{MIG-GT} &0.0656&\underline{0.1009}&0.0353&0.0442&0.0742&0.1085& \underline{0.0403}&0.0491&0.0619&0.0904&0.0338&0.0410\\
        \textbf{BeFA} & 0.0555 & 0.0884 & 0.0299 & 0.0383 & 0.0649 & 0.0985 & 0.0346 & 0.0432 & 0.0568 & 0.0857 &
        0.0307 & 0.0381 \\
    \midrule   \textbf{AMUR}&\textbf{0.0692}*&\textbf{0.1025}*&\textbf{0.0371}*&\textbf{0.0457}*&\textbf{0.0791}*&	\textbf{0.1182}*&\textbf{0.0439}*&\textbf{0.0540}*&\textbf{0.0686}*&\textbf{0.0997}*&\textbf{0.0375}*&\textbf{0.0454}* \\
    \textbf{\%Improv.}&4.69\%&1.56\%&4.21\%&2.24\%&5.33\%&3.96\%&8.93\%&7.78\%&3.78\%&2.26\%&4.17\%&3.65\% \\
    \bottomrule
    \end{tabular}
     }
\end{table*}

Table~\ref{tab:comparison} reports the overall performance on three real-world multimodal recommendation datasets. 
AMUR achieves the best results across all datasets and evaluation metrics, showing the effectiveness of information-guided selective modality-interest alignment. 
We summarize the key observations as follows:

\begin{itemize}

\item \textbf{Overall effectiveness.}
AMUR consistently outperforms representative baselines from different categories, including MF-based methods, invariant/causal learning methods, graph-based methods, SSL/alignment-based methods, diffusion-based methods, and Transformer/MLLM-based methods. 
Compared with the strongest baseline on each metric, AMUR improves Recall@20 by 1.56\%, 3.96\%, and 2.26\% on Baby, Sports, and Clothing, respectively, and improves NDCG@20 by 2.24\%, 7.78\%, and 3.65\%. 
These consistent improvements indicate that selectively aligning modality information with user interests is beneficial for multimodal recommendation.

\item \textbf{Advantages over recent denoising and alignment baselines.}
AMUR also outperforms strong recent baselines such as DA-MRS, MENTOR, DiffMM, CCDRec, TMLP, MIG-GT, and BeFA. 
This shows that existing denoising, alignment, diffusion-based, or Transformer-based designs may still suffer when modality information is not selectively aligned with user interests. 
By combining selective modality-interest refinement and selective cross-modal shared-interest alignment, AMUR better emphasizes preference-aligned modality semantics while preserving useful modality-specific information.

\end{itemize}

\subsection{Ablation Study}
\label{sect:ablation}

To evaluate the contribution of each component in AMUR, we construct four ablated variants:

\begin{itemize}
    \item \textbf{w/o BR}. 
    This variant removes the behavior-calibrated graph refinement module. The modality graph is directly used without being guided by behavioral co-occurrence signals.

    \item \textbf{w/o CIE}. 
    This variant removes the interest-aware contrastive learning objective, so modality representations are not explicitly pulled toward the collaborative interests of interacted users.

    \item \textbf{w/o DR}. 
    This variant removes the shared-subspace discrepancy regularization, weakening the constraint that reduces unnecessary discrepancy between selected visual and textual shared-interest representations.

    \item \textbf{w/o CE}. 
    This variant removes the shared-subspace consistency maximization objective, thus discarding the cross-modal contrastive signal for enhancing shared interest-related semantics.
\end{itemize}

\begin{figure}[t]
    \centering
    \includegraphics[width=1\linewidth]{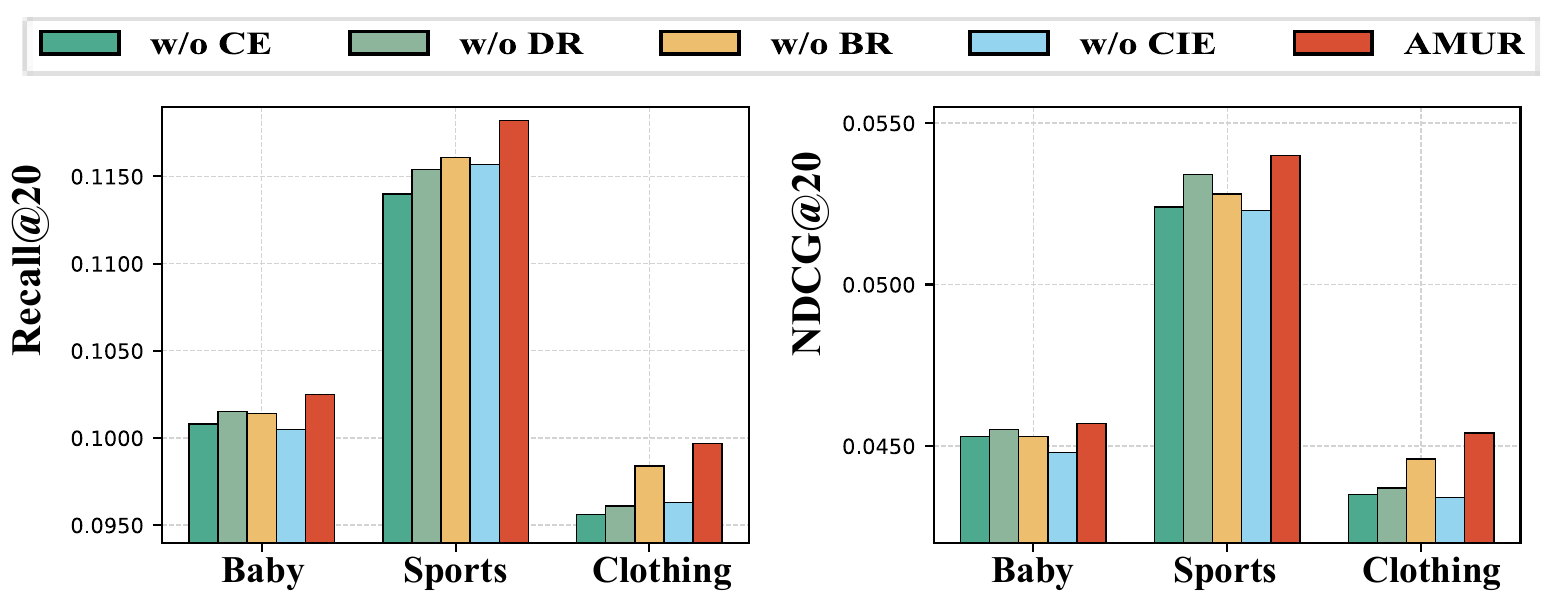}
    \caption{Ablation results of different components.}
    \Description{Bar charts of ablation results across different datasets and evaluation metrics, comparing the full AMUR model with variants that remove individual components.}
    \label{fig:ablation}
\end{figure}

The results are shown in Figure~\ref{fig:ablation}. We have the following observations. 
\textbf{First}, removing any component leads to performance degradation across datasets, demonstrating that all modules contribute to the final recommendation performance. Specifically, BR helps refine modality structures with behavioral signals, CIE encourages modality representations to be aligned with user interests, DR reduces unnecessary discrepancy in the selected shared-interest subspace, and CE enhances discriminative shared semantics across modalities.

\textbf{Second}, the variants w/o CE and w/o DR generally suffer clear performance drops, indicating the importance of selective cross-modal shared-interest alignment. This shows that, after modality representations are refined toward user interests, exploiting shared interest-related semantics across modalities can further improve recommendation quality.

\textbf{Third}, removing BR or CIE also hurts performance, especially on datasets where modality signals are more diverse or noisy. This verifies the necessity of selective modality-interest refinement: behavioral graph refinement helps reduce the influence of less aligned modality relations, while interest-aware contrastive learning further guides modality representations toward user preferences.

\begin{table}[t]
\centering
\caption{Effectiveness of behavior-guided graph refinement and shared-interest subspace selection.}
\label{tab:design_analysis}
\resizebox{\linewidth}{!}{
\begin{tabular}{lcccccc}
\toprule
\multirow{2}{*}{Variants} 
& \multicolumn{2}{c}{Baby} 
& \multicolumn{2}{c}{Sports} 
& \multicolumn{2}{c}{Clothing} \\
\cmidrule(lr){2-3} \cmidrule(lr){4-5} \cmidrule(lr){6-7}
& R@20 & N@20 
& R@20 & N@20 
& R@20 & N@20 \\
\midrule
MM emb 
& 0.0995 & 0.0443
& 0.1104 & 0.0490
& 0.0921 & 0.0409 \\
Full space 
& 0.0992 & 0.0441
& 0.1143 & 0.0520
& 0.0974 & 0.0437 \\
AMUR
& \textbf{0.1025} & \textbf{0.0457}
& \textbf{0.1182} & \textbf{0.0540}
& \textbf{0.0997} & \textbf{0.0454} \\
\bottomrule
\end{tabular}
}
\end{table}

Table~\ref{tab:design_analysis} further validates two key designs of AMUR.
The \textbf{MM emb} variant replaces behavior-guided graph refinement with
multimodal embedding-based edge estimation, while \textbf{Full space} removes
shared-interest subspace selection and aligns the full modality
representations. AMUR consistently outperforms both variants across all
datasets. This demonstrates that behavioral signals are more reliable than
raw multimodal similarity for preference-aware graph refinement, and that
selective shared-subspace alignment is more effective than full-space
cross-modal alignment.

\subsection{In-depth Analysis}
\label{indepth}

To further understand how AMUR improves modality-interest alignment, we conduct several in-depth analyses, including visualization, feature-importance study, information-theoretic trend analysis, robustness analysis, and hyperparameter study.

% \begin{figure}[t]
%     \centering
%     \includegraphics[width=1\linewidth]{figures/vis.pdf}
%     \caption{Visualization of modality-interest alignment.}
%     \label{fig:vis}
% \end{figure}

\begin{figure}[t]
    \centering
    \includegraphics[width=\linewidth]{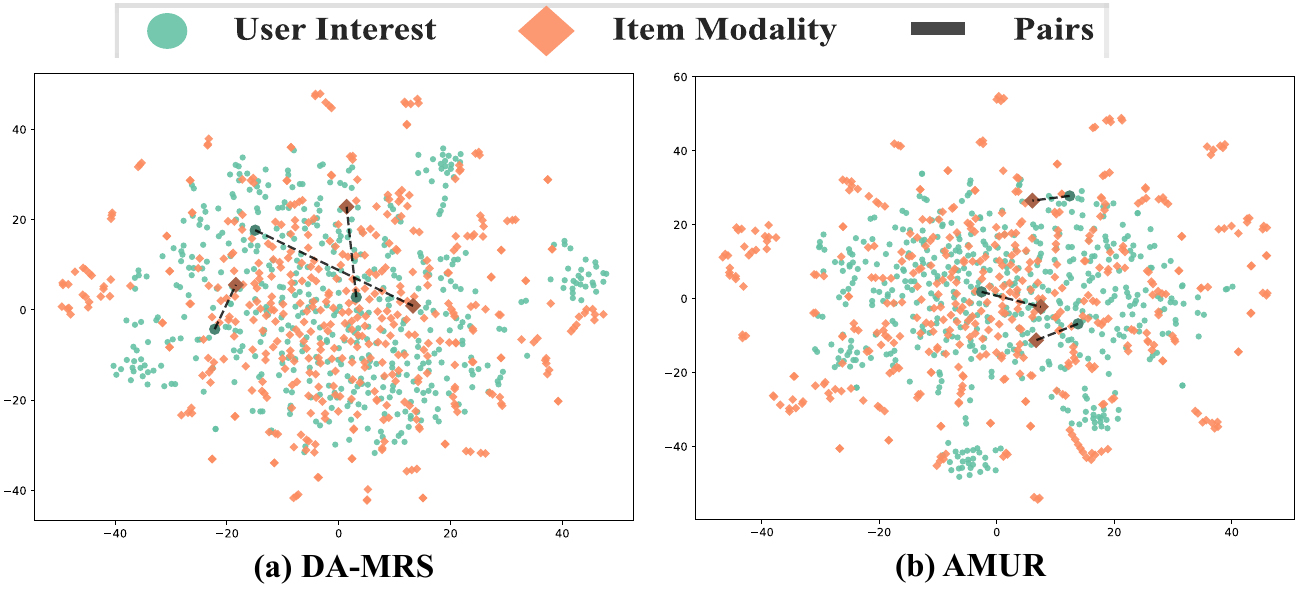}
    \caption{Visualization of modality-interest alignment.}
    \Description{Two scatter plots comparing modality-interest alignment between DA-MRS and AMUR. Green circles represent user interests, orange diamonds represent item modality representations, and gray line segments connect paired representations. Compared with DA-MRS, AMUR shows shorter distances between paired representations, indicating tighter modality-interest alignment.}
    \label{fig:vis}
\end{figure}

\textbf{Visualization of Modality--Interest Alignment.}
We visualize the t-SNE projections of item modality representations and their corresponding user representations in Figure~\ref{fig:vis}. 
Compared with DA-MRS, AMUR produces more compact clusters between users and their interacted items, indicating that the learned modality representations are more closely related to user interests.

% \begin{figure}[t]
%     \centering
%     \includegraphics[width=1\linewidth]{figures/residualtsne.pdf}
%     \caption{Visualization of aligned and residual modality representations.}
%     \label{fig:restsne}
% \end{figure}

\begin{figure}[t]
    \centering
    \includegraphics[width=\linewidth]{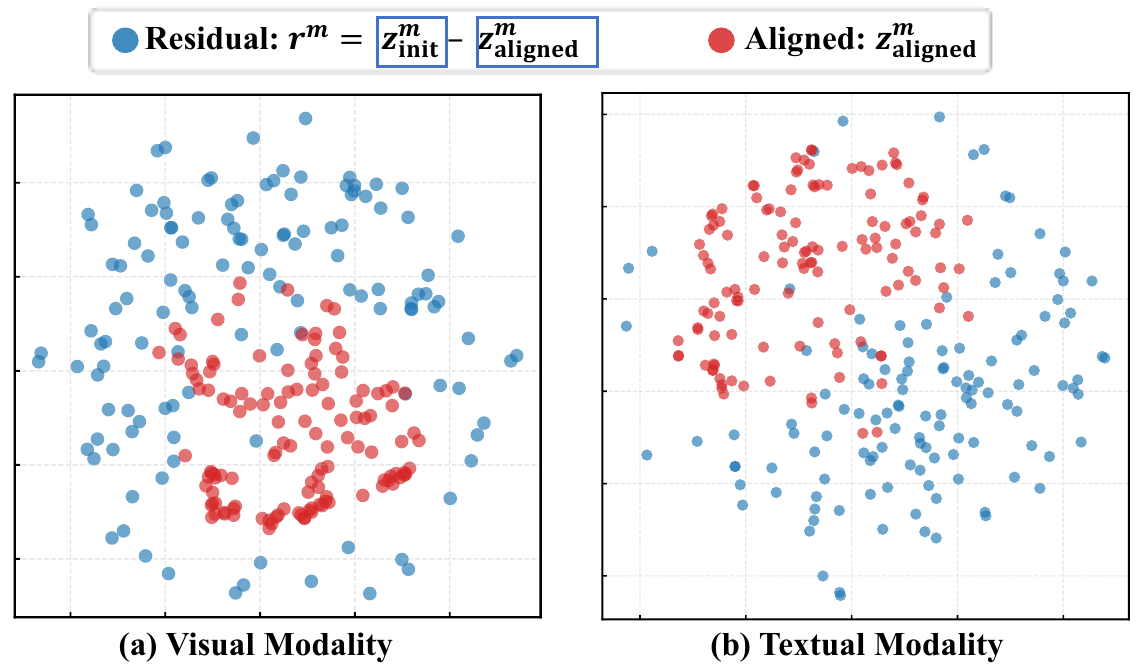}
    \caption{Visualization of aligned and residual modality representations.}
    \Description{Two scatter plots visualize aligned and residual modality representations for visual and textual modalities. Red points denote aligned representations, while blue points denote residual representations obtained by subtracting the aligned component from the initial modality representation. The aligned and residual representations form distinguishable distributions in both modalities, illustrating the decomposition of modality information into aligned and residual components.}
    \label{fig:restsne}
\end{figure}

\textbf{Visualization of Aligned and Residual Representations.}
We further compare the aligned modality representations with the residual representations, where the residual is computed as
\[
\boldsymbol{r}^m = \boldsymbol{z}_{\mathrm{init}}^m - \boldsymbol{z}_{\mathrm{aligned}}^m .
\]
As shown in Figure~\ref{fig:restsne}, the aligned representations form more compact clusters, while the residual representations are more dispersed. 
This suggests that AMUR concentrates user-aligned modality semantics into the aligned representations and reduces the influence of less structured residual signals.

\begin{figure}[t]
    \centering
    \includegraphics[width=1\linewidth]{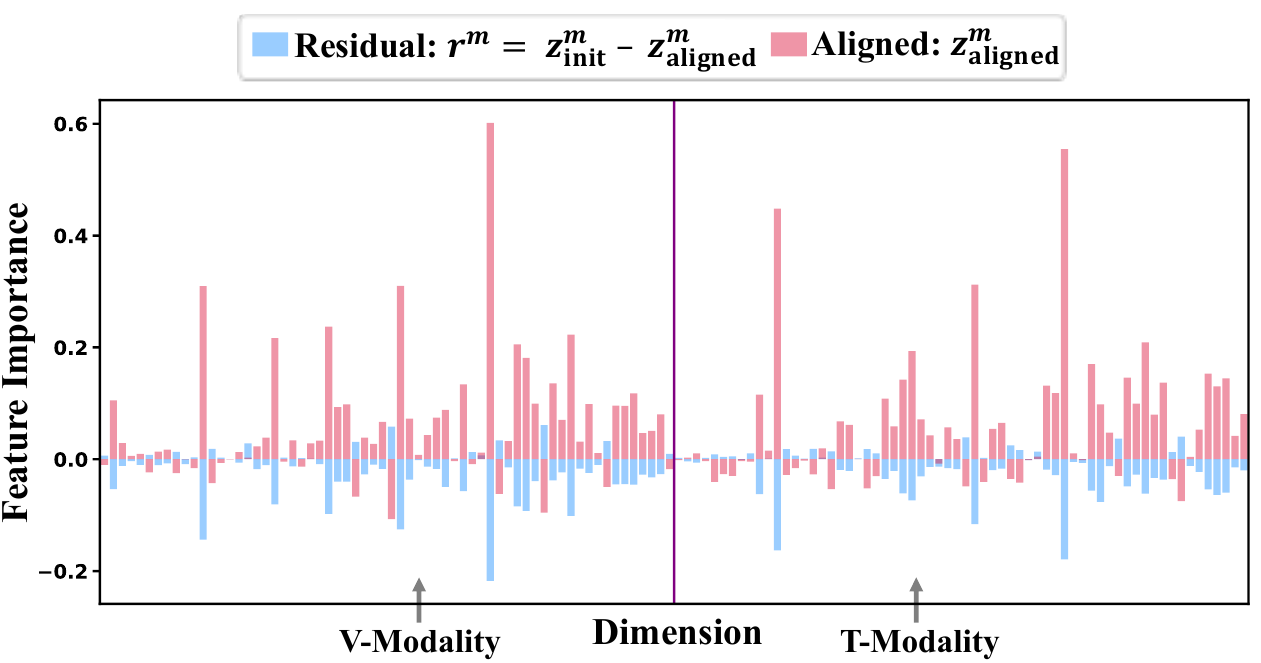}
    \caption{Feature-importance analysis of aligned and residual modality representations.}
    \Description{Bar charts of feature importance for visual and textual modality representations. Pink bars denote the aligned representations, while blue bars denote the residual representations obtained by subtracting the aligned component from the initial modality representation. The two components exhibit clearly different importance patterns across feature dimensions in both modalities.}
    \label{fig:featimprt}
\end{figure}

\textbf{Feature-Importance Analysis.}
For each dimension \(k\), we mask it as zero and measure the resulting decrease in the user--item score:
\[
\Delta_k = \mathrm{Score}_{\mathrm{full}} - \mathrm{Score}_{\mathrm{without\ dim}\ k}.
\]
As shown in Figure~\ref{fig:featimprt}, the aligned representations generally show higher importance values than the residual representations. 
This indicates that AMUR preserves more prediction-useful information in the aligned representations, while the residual part contains more weakly useful or potentially distracting signals.

\begin{figure*}[t]
    \centering
    \includegraphics[width=0.8\linewidth]{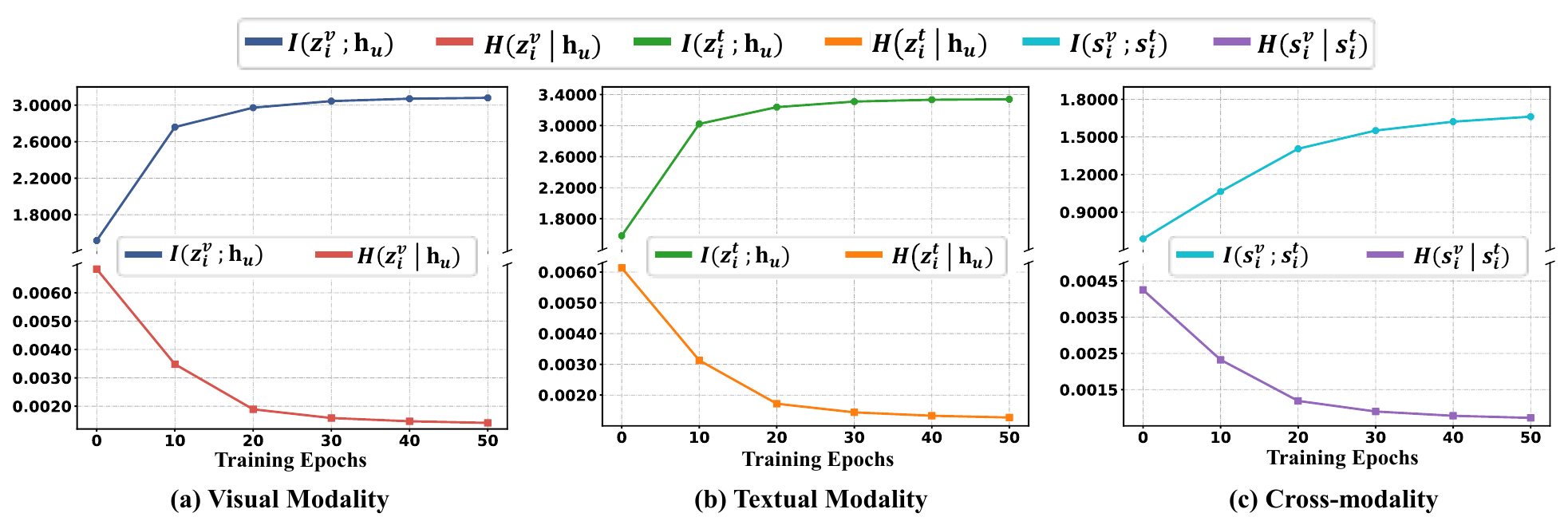}
    \caption{Estimated mutual information and conditional entropy during training.}
    \Description{Three line plots show the evolution of estimated mutual information and conditional entropy over training epochs for visual modality, textual modality, and cross-modality representations. In all three cases, mutual information increases and gradually stabilizes as training proceeds, while the corresponding conditional entropy decreases, indicating progressively stronger information alignment.}
    \label{fig:infoterm}
\end{figure*}

\textbf{Information-Theoretic Trend Analysis.}
We estimate mutual information with MINE~\cite{belghazi2018mutual} and conditional entropy with a neural variational estimator. 
As shown in Figure~\ref{fig:infoterm}, \(I(\boldsymbol{z}_i^v;\boldsymbol{\mathrm{h}}_u)\) and \(I(\boldsymbol{z}_i^t;\boldsymbol{\mathrm{h}}_u)\) gradually increase, while \(H(\boldsymbol{z}_i^v|\boldsymbol{\mathrm{h}}_u)\) and \(H(\boldsymbol{z}_i^t|\boldsymbol{\mathrm{h}}_u)\) decrease. 
For cross-modal shared-interest alignment, \(I(\boldsymbol{s}_i^v;\boldsymbol{s}_i^t)\) increases and \(H(\boldsymbol{s}_i^v|\boldsymbol{s}_i^t)\) decreases. 
These trends are consistent with the information-guided motivation of AMUR.

\begin{figure}[t]
    \centering
    \includegraphics[width=1\linewidth]{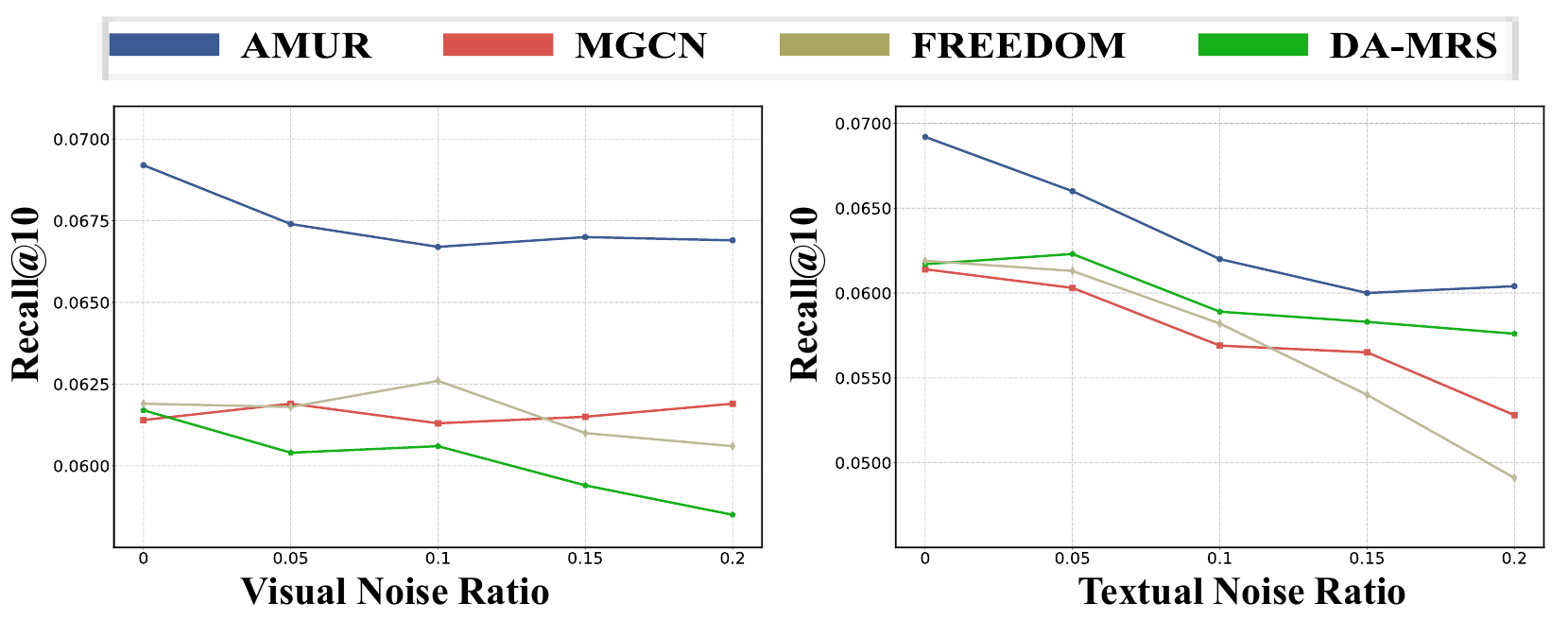}
    \caption{Performance under corrupted modality content.}
    \Description{Two line plots show Recall@10 of AMUR, MGCN, FREEDOM, and DA-MRS under increasing visual and textual noise ratios. AMUR remains relatively stable across different corruption levels, while the baseline methods show varying degrees of performance decline.}
    \label{fig:denoising}
\end{figure}

\textbf{Robustness to Corrupted Modality Content.}
We randomly corrupt a portion of item modality content by replacing it with the modality content of another item. 
In Figure~\ref{fig:denoising}, AMUR consistently outperforms the compared baselines under different corruption ratios, demonstrating its robustness to noisy modality signals.

\begin{figure}[t]
    \centering
    \includegraphics[width=1\linewidth]{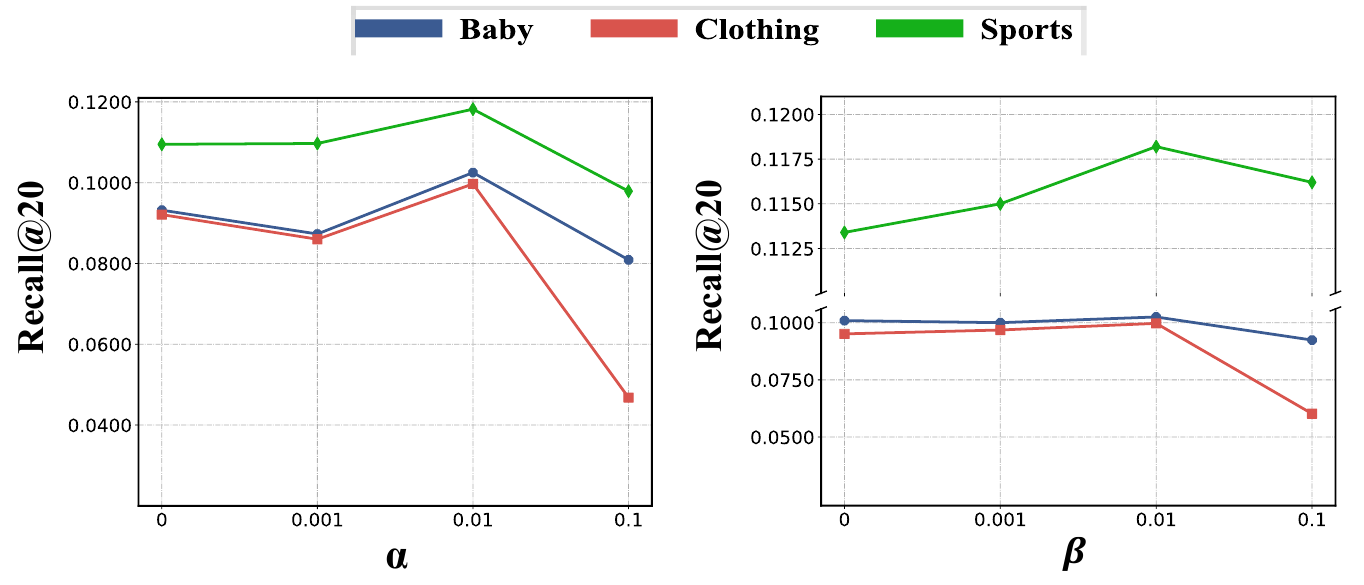}
    \caption{Hyperparameter study on coefficients \(\alpha\) and \(\beta\).}
    \Description{Two line plots show Recall@20 on the Baby, Clothing, and Sports datasets under different values of the hyperparameters \(\alpha\) and \(\beta\). The three datasets exhibit different sensitivity to these coefficients, with performance varying more noticeably at larger values.}
    \label{fig:hyper}
\end{figure}

\textbf{Hyperparameter Analysis.}
We study the influence of \(\alpha\) and \(\beta\), which control selective modality-interest refinement and selective cross-modal shared-interest alignment, respectively. 
As shown in Figure~\ref{fig:hyper}, performance generally improves as each coefficient increases from 0 and reaches the best result around 0.01. 
Further increasing the coefficient may degrade performance, indicating the need to balance recommendation supervision and auxiliary alignment objectives.

\section{Related Work}

\subsection{Multimodal Recommendation}

Multimodal recommendation aims to improve user preference modeling by leveraging rich modality content such as images and text. Early studies extend matrix factorization with modality features, such as VBPR~\cite{he2016vbpr} and CKE~\cite{zhang2016collaborative}. With the development of graph neural networks, graph-based methods have been widely explored for modeling collaborative and multimodal signals~\cite{wei2019mmgcn, wang2021dualgnn, zhou2023tale, su2024soil, huang2025beyond, hu2025modality, jeong2025melon, xu2025best}. Representative methods include MMGCN~\cite{wei2019mmgcn}, which propagates modality-specific representations over user--item graphs, and LGMRec~\cite{guo2024lgmrec}, which models local and global user interests with multimodal graph learning. 

Self-supervised methods further improve multimodal representation learning through contrastive or reconstruction objectives~\cite{tao2022self, zhou2023bootstrap, yang2024multimodal, yi2022multi, li2024msi}, such as MGCN~\cite{yu2023multi} and BM3~\cite{zhou2023bootstrap}. Recently, diffusion-based methods have also been introduced into multimodal recommendation~\cite{jiang2024diffmm, cui2025multi, ma2025generating, he2025flip}, such as DiffMM~\cite{jiang2024diffmm}, MCDRec~\cite{ma2024multimodal}, and LD4MRec~\cite{yu2023ld4mrec}.

Although these methods have achieved promising performance, they mainly focus on how to incorporate modality information into recommendation models, while paying less attention to whether the introduced modality signals are well aligned with user interests. In contrast, our work focuses on selective modality-interest alignment, aiming to emphasize modality semantics that better match user preferences while reducing the influence of less aligned signals.

\subsection{Denoising and Aligning Modality Information for Recommendation}

Noisy and misaligned modality information is an important challenge in multimodal recommendation. Existing methods address this issue from different perspectives, including modality denoising~\cite{yu2023multi, ong2025spectrum, yang2025fitmm}, graph refinement~\cite{zhang2021mining, wei2020graph, chen2020iterative, sun2022graph, zhang2022latent}, diffusion-based representation refinement~\cite{ma2024multimodal, yu2023ld4mrec}, invariant representation learning~\cite{du2022invariant}, and contrastive or self-supervised alignment~\cite{liu2024aligning, xv2024improving, xu2025mentor}.

However, existing denoising and alignment strategies are often implicit or heuristic. They usually lack a clear objective for selectively aligning modality information with user interests, making it difficult to explain why certain modality signals should be emphasized or weakened. Moreover, many cross-modal alignment strategies encourage consistency across full modality representations, which may overlook useful modality-specific complementary information. Different from these methods, AMUR uses an information-guided view to refine modality information toward user interests and align only the selected shared-interest subspace across modalities.

\subsection{Runtime Analysis}
\label{sect:runtime}

We further evaluate the training efficiency of AMUR on the Sports dataset using an NVIDIA GeForce RTX 3060. 
Table~\ref{tab:runtime} reports the number of epochs, total training cost, and Recall@20 of different models.
As shown in Table~\ref{tab:runtime}, AMUR achieves the best performance with a moderate training cost. 
Although MGCN is slightly faster, its performance is clearly lower than AMUR. 
Compared with strong baselines such as DA-MRS, MIG-GT, and TMLP, AMUR achieves higher Recall@20 while requiring less training time. 
These results show that AMUR provides a good trade-off between effectiveness and efficiency, indicating that the proposed selective alignment objectives do not introduce excessive training overhead.

\begin{table}[t]
\centering
\caption{Runtime analysis on the Sports dataset.}
\label{tab:runtime}
\resizebox{0.7\linewidth}{!}{
\begin{tabular}{lccc}
\toprule
\textbf{Model} & \textbf{Epochs} & \textbf{Training Cost} & \textbf{R@20} \\
\midrule
BM3      & 246 & 28.4 min & 0.0981 \\
MGCN     & 76  & \textbf{11.4 min} & 0.1105 \\
LGMRec   & 60  & 15.7 min & 0.1090 \\
DA-MRS   & 63  & 43.8 min & 0.1073 \\
MIG-GT   & 300 & 18.5 min & 0.1085 \\
TMLP     & 153 & 36.3 min & \underline{0.1137} \\
\midrule
AMUR     & 67  & \underline{12.9 min} & \textbf{0.1187} \\
\bottomrule
\end{tabular}
}
\end{table}

\section{Conclusions and Future Work}

In this work, we proposed AMUR, an information-guided selective modality-interest alignment framework for multimodal recommendation. AMUR refines modality information toward user interests and further aligns shared interest-related semantics across modalities, while preserving modality-specific complementary information. Experiments on three real-world datasets demonstrate the effectiveness of AMUR, and further analyses verify the contribution of its key components.

In the future, we will explore more fine-grained preference-aligned modality modeling and extend the proposed selective alignment idea to broader recommendation scenarios.

\section{GenAI Usage Disclosure}
Generative AI tools were used solely for language polishing and improving the clarity of the manuscript. All research ideas, methodological designs, experiments, analyses, and conclusions were developed and verified by the authors.

\begin{acks}
This research is supported in part by National Science Foundation of China (No. 62472277, No. 62072304), Shanghai Municipal Science and Technology Commission (No. 21511104700), the Shanghai East Talents Program (2023-177).
\end{acks}

\bibliographystyle{ACM-Reference-Format}
\balance
\bibliography{sample-base}
\end{document}